# Visualization of Acoustic Power Flow in Suspended Thin-film Lithium Niobate Phononic Devices

Daehun Lee[1], Shawn Meyer[1], Songbin Gong[3], Ruochen Lu*[2], Keji Lai*[1]

[1] Department of Physics, University of Texas at Austin, Austin TX 78712, USA

[2] Department of Electrical and Computer Engineering, University of Texas at Austin, Austin TX 78712, USA

[3] Department of Electrical and Computer Engineering, University of Illinois at Urbana-Champaign, Urbana IL, 61801, USA

* E-mails: ruochen@utexas.edu, kejilai@physics.utexas.edu

## Abstract

We report direct visualization of gigahertz-frequency acoustic waves in lithium niobate phononic circuits. Primary propagation parameters, such as the power flow angle (PFA) and propagation loss (PL), are measured by transmission-mode microwave impedance microscopy (TMIM). Using fast Fourier transform, we can separately analyze forward and backward propagating waves and quantitatively evaluate the propagation loss. Our work provides insightful information on the propagation, diffraction, and attenuation in piezoelectric thin films, which is highly desirable for designing and optimizing phononic devices for microwave signal processing.

Acoustic-wave devices have been the primary radio frequency (RF) mobile front-end filtering solution in the telecommunication industry [1]. Their commercial success is mostly enabled by shorter wavelength and lower damping of the acoustic wave than the electromagnetic counterpart, resulting in compact signal processing elements with low loss and good frequency selectivity [2]. More recently, as 5G communication and beyond moves toward higher frequencies (> 3 GHz) and wider bandwidths (> 10%) [3], it is not trivial for conventional surface acoustic wave (SAW) [4] and bulk acoustic wave (BAW) [5] platforms to meet the new specifications, as their performance is fundamentally limited by the moderate electromechanical coupling ($K^2$) and quality factor ($Q$) at higher frequencies [6]. New acoustic platforms with better frequency scalability and simultaneously high $K^2$ and $Q$ are thus highly sought after.

In the past decade, Lamb-wave devices in lithium niobate ($LiNbO_3$) single-crystal thin films have been identified as promising candidates for GHz wideband acoustic signal processing [7]. Thanks to the advance of thin-film transfer technology [8], various acoustic elements have been implemented in suspended $LiNbO_3$ as mm-wave resonators [9], filters [10, 11], delay lines [12], and amplifiers [13]. Despite the enhanced performance promised by the new platform, it is more challenging to design and optimize GHz acoustic devices in suspended thin-film $LiNbO_3$ than in conventional piezoelectric platforms. First of all, acoustic waves in thin-film $LiNbO_3$ are highly anisotropic, causing deviation in direction between phase velocity ($v_{ph}$) and group velocity ($v_g$), i.e., slanted power flow described by a non-zero power flow angle (PFA) [14]. Without considering PFA, resonant devices with low $Q$ and propagation devices with energy launching off the receiving transducers are inevitable. Secondly, the propagation loss (PL), diffraction, and reflection of GHz acoustic waves in thin-film $LiNbO_3$ are understudied. Optimizing device performance, e.g., minimizing loss, is thus difficult. Thirdly, the highly piezoelectric substrates tend to generate spurious modes, i.e., unwanted acoustic vibration near the main tone, via electrical routing and parasitics. It is difficult to identify the origin of such effects merely from electrical measurement. Last but not least, the characterization of acoustic components is limited to terminal-to-terminal electrical response. Researchers have to mostly rely on numerical modeling to understand the GHz acoustic vibration, with little experimental validation on the actual displacement at the nanoscale.

One effective approach for designing $LiNbO_3$ phononic devices is to directly visualize the acoustic vibration. If available, such measurement would facilitate performance optimization in conventional functions and enable novel applications, e.g., capturing energy leakage in acoustic resonators or mode conversion in phononic circuits. It is also applicable to emerging applications such as topological phononic crystals and acoustic metamaterials. To achieve this goal, many attempts, including optical reflectometry [15, 16] and interferometry [17, 18], stroboscopic X-ray imaging [19, 20], and scanning electron microscopy [21, 22], among others, have been made to probe the acoustic displacement fields. However, none of these techniques can simultaneously achieve sub-100 nm spatial resolution and > 1 GHz operation frequency, which are crucial for technologically relevant devices. A new method that is capable of investigating the nanoscale GHz acoustic phenomena is therefore highly desirable for designing and optimizing advanced phononic circuits.

In this work, we report the visualization of 1 GHz fundamental symmetric mode ($S_0$ mode) in suspended thin-film $LiNbO_3$ by transmission-mode microwave impedance microscopy (TMIM) [23-25]. The power flow angle (PFA) is directly measured from the TMIM images and compared to the finite-element simulation. Using fast Fourier transform (FFT), the forward and backward propagation waves are quantitatively analyzed and the propagation loss calculated. Our work provides insightful information on energy flow inside acoustic structures and suggests a new approach to characterize phononic devices.

The suspended phononic devices in this work were fabricated on 800 nm X-cut $LiNbO_3$ thin film transferred onto Si carrier chips (NGK Insulators, Ltd). The sputtered electrodes (135 nm thick) were defined by e-beam lithography. A $SiO_2$ layer (2 μm in thickness) was deposited using plasma-enhanced chemical vapor deposition (PECVD) as the hard mask and patterned by fluorine-based reactive ion etching (RIE). The release windows in the $LiNbO_3$ film were subsequently etched using chlorine-based inductively coupled plasma reactive ion etching (ICP-RIE). The remaining $SiO_2$ was removed by low-power ICP. The devices were then released by isotropic $XeF_2$-based silicon etching. As shown in Fig. 1a, the acoustic waveguide (width $w = 150$ μm and length $L = 700$ μm) with 4 ports of single-phase unidirectional transducers (SPUDTs) is oriented 15° away from the y axis. [26]

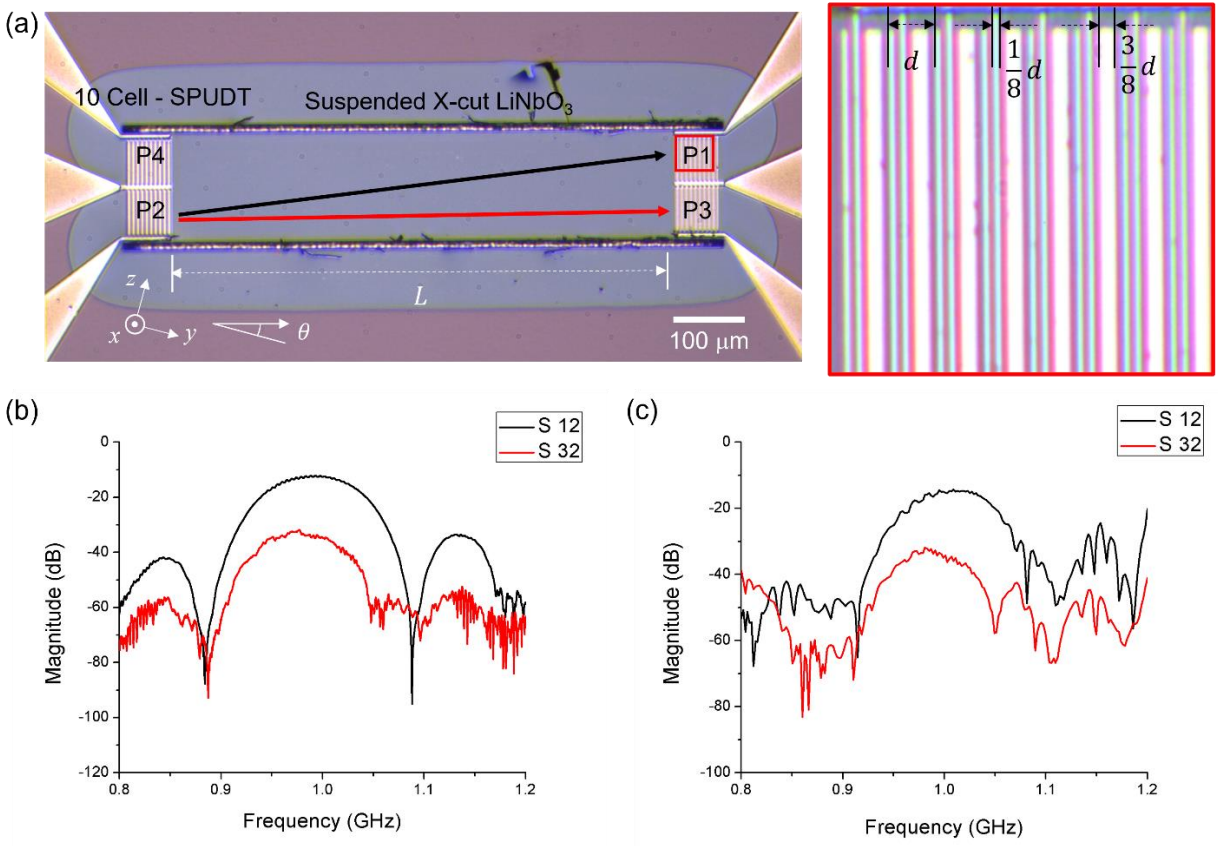

FIG. 1. **(a)** Optical image of the suspended X-cut LiNbO$_3$ waveguide and four ports of 10-cell single-phase unidirectional transducers (SPUDTs). The waveguide direction is tilted by θ = 15° from the y-axis. The red and black arrows indicate the power transmission from Port 2 to Port 3 and Port 1, respectively. The inset on the right shows a magnified image of the boxed region. One unit cell of the SPUDT consists of three fingers, one ($d$/8 in width, where $d$ is the wavelength) connected to the signal electrode and two ($d$/8, and 3$d$/8 in widths) connected to the ground electrode. **(b)** $S_{12}$ and $S_{32}$ spectra of the device measured by a VNA, showing the acoustic passband at $f$ = 1.00 GHz. Note that $S_{12}$ is significantly higher than $S_{32}$ in the passband. **(c)** Finite-element simulation results of $S_{12}$ and $S_{32}$ spectral.

Electrical characterization of the LiNbO$_3$ waveguide test structure was first carried out with a vector network analyzer (VNA). Here the acoustic wave is launched from Port 2 and received at Ports 1 and 3. The measured transmission coefficients $S_{12}$ and $S_{32}$ are plotted in Fig. 1b. Inside the S$_0$ passband centered at $f$ = 1.00 GHz, it is clear that $S_{12}$ is significantly higher than $S_{32}$ by ~ 20 dB, indicative of the slanted power flow. In order to understand the power transmission, we also modeled the SPUDTs and acoustic waveguide by finite-element analysis (FEA). The simulated $S_{12}$ and $S_{32}$ in Fig. 1c are in good agreement with the measured results in Fig. 1b.

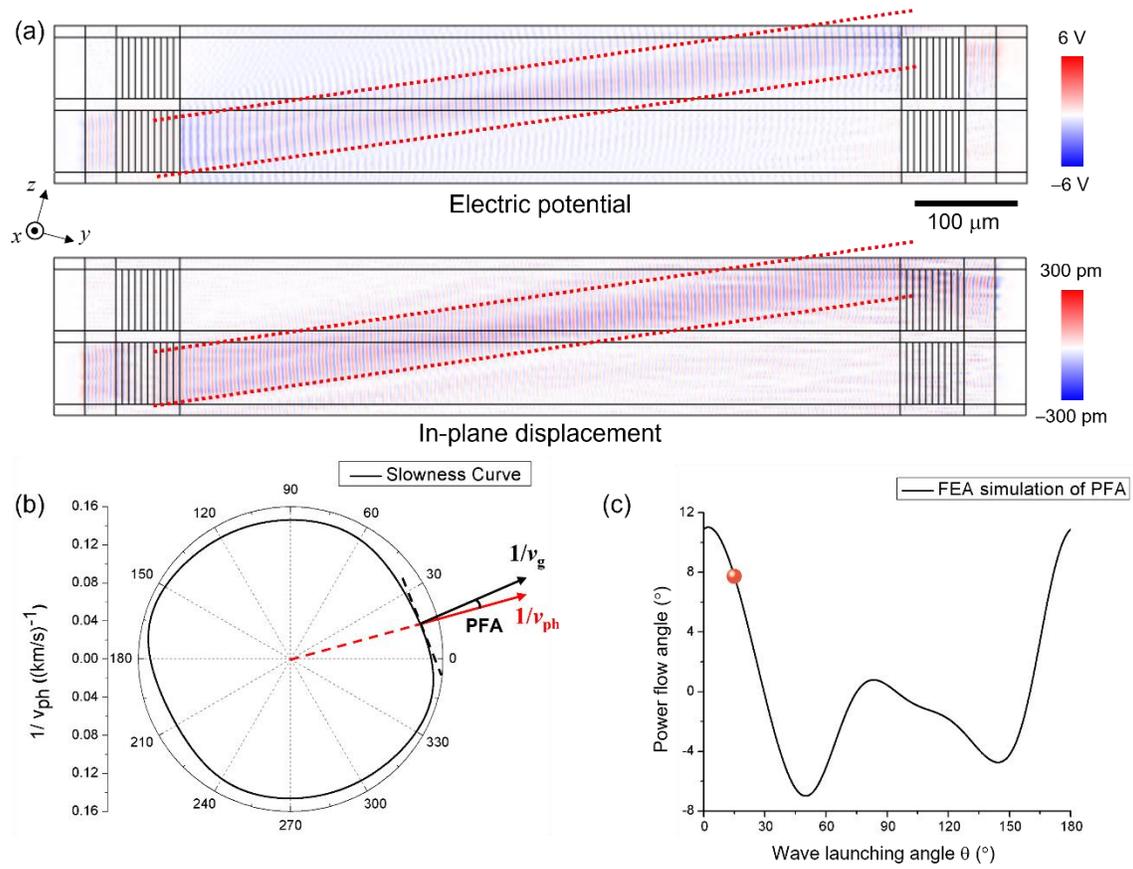

FIG. 2. **(a)** FEA of electric potential (upper panel) and in-plane displacement along the waveguide direction (lower panel) of the LiNbO$_3$ thin-film waveguide device under 3 dBm excitation power. **(b)** FEA simulated slowness curve of S$_0$ (thick black curve) in unmetallized LiNbO$_3$ thin film along different angles to +y-axis (outer circle). The acoustic wave in this work is launched at θ = 15° (red dashed line and arrow). The direction of the group velocity is normal to the slowness curve (black arrow). The angle between the two vectors defines the power flow angle (PFA). **(c)** Simulated PFA as a function of the wave launching angle. The orange dot denotes the PFA at θ = 15°.

The simulated surface potential and total in-plane displacement field on the LiNbO$_3$ device are shown in Fig. 2a. The results clearly demonstrate that the slanted acoustic wave propagation, which stems from the difference in direction between phase velocity ($v_{ph}$) and group velocity ($v_g$) [14]. Fig. 2b plots the S$_0$ slowness curve, i.e., 1/$v_{ph}$ along launching angles from 0 to 2. The PFA is then determined by the angle between the phase velocity (along the wave launching by SPUDT) and the group velocity (along the normal line of the slowness curve). Fig. 2c shows the simulated

PFA as a function of the wave launching angle. At θ = 15°, the simulation indicates that PFA is ~ 7.7°, which is consistent with previous reports [27, 28].

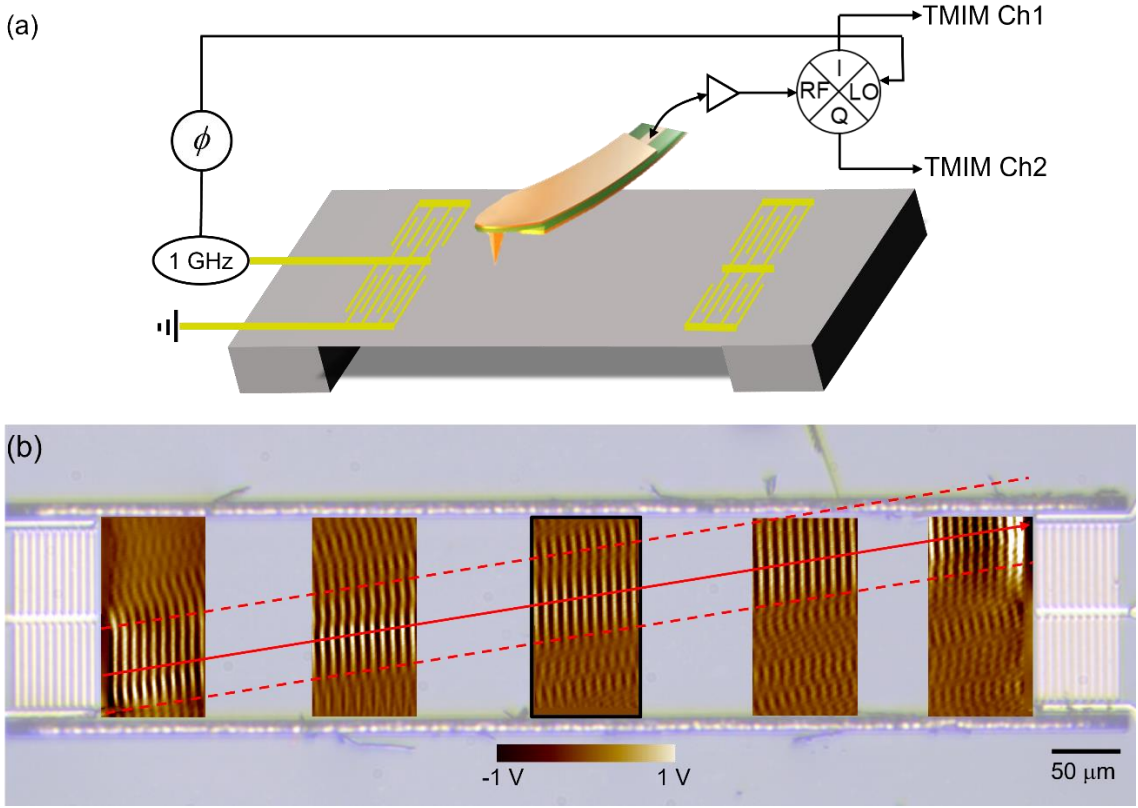

FIG. 3. **(a)** Schematic of the TMIM imaging on a suspended waveguide. **(b)** TMIM images overlaid on top of an optical picture of the device. The red arrow and dashed lines denote the propagation direction and approximate boundaries of the acoustic wave.

Visualization of the Lamb wave on thin-film LiNbO$_3$ waveguide was carried out by TMIM, an atomic-force-microscopy (AFM)-based technique with sub-100nm spatial resolution [23-25]. Fig. 3a shows the schematic of the experiment, where the acoustic wave is launched by the emitter SPUDT and the induced piezoelectric potential at 1.00 GHz is detected by the tip. The signal is then amplified and demodulated by an in-phase/quadrature (I/Q) mixer, which converts time-varying acoustic signals to time-independent impedance images. The complex-valued TMIM signal $V_{Ch1} + i * V_{Ch2}$ thus provides a phase-sensitive measurement on the local displacement field of the Lamb wave. Fig. 3b shows images from one of the two TMIM channels taken at various locations on the device. The prominent Lamb wave launched from Port 2 is planewave-like with wavefront parallel to the emitter SPUDT. The acoustic power flow, on the other hand, clearly deviates from the wave launching direction. By tracking the maximum intensity in the TMIM images along the LiNbO$_3$ waveguide (red arrow), we can estimate a PFA ~ 8° for this device, which matches well with the simulated result above. It should be noted that, in addition to the primary wave bounded by the dashed lines in Fig. 3b, scattered waves due to waveguide boundaries and interference pattern due to multiple reflections are observed in the TMIM images. These imperfections are sample-dependent and difficult to be simulated by FEA. Careful analysis of these secondary features could help optimize device performance in future studies.

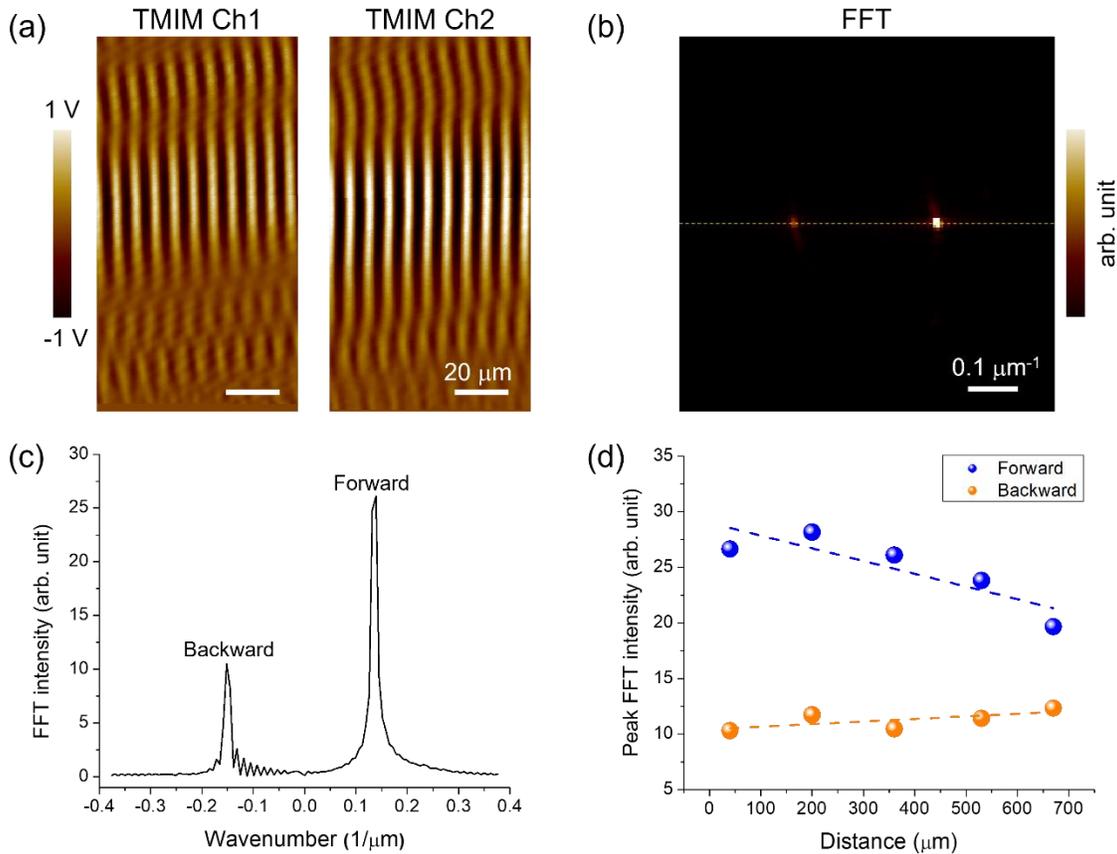

FIG. 4. **(a)** Simultaneously demodulated TMIM Ch1 and Ch2 images near the center of the testbed device. **(b)** FFT amplitude map of the TMIM data in **a**. **(c)** Line profile across the yellow dashed line in **b**. **(d)** Peak FFT intensity of TMIM results as a function of distance from the wave launching port. The blue and orange dots represent the forward and backward propagation, respectively. Dashed lines are linear fits to the data.

In addition to imaging acoustic waves, the TMIM also allows us to perform quantitative analysis of the power flow in LiNbO$_3$ waveguides. Specifically, the momentum-space information can be obtained by taking FFT [25] of the combined signal $V_{Ch1} + i * V_{Ch2}$. Fig. 4a displays the TMIM images from the two orthogonal channels at the center of the suspended device (outlined in black in Fig. 3b). The corresponding FFT amplitude map near zero wavenumber is shown in Fig. 4b, where two high-intensity spots are clearly seen along the waveguide direction. In Fig. 4c, we plot the line profile across the dashed line in Fig. 4b. The peaks on the left and right correspond to the right-moving and left-moving Lamb waves, respectively. Since the FFT amplitude is directly proportional to the magnitude of the waves, we can estimate that, at this location, the power of the forward-propagating wave is ~ 7 times higher than that of the backward one from the ratio between these two intensity peaks. By plotting the peak height as a function of the position (Fig. 4d), we can calculate PLs of ~ 4.0 dB/mm for the forward wave and ~ 1.8 dB/mm for the backward wave, which are comparable to that reported in suspended LiNbO$_3$ thin films [29]. The difference in PL between wave propagating in opposite directions requires further investigations. Nevertheless, we emphasize that such phase-sensitive information is not readily available from other spatially resolved methods that mainly probe the surface oscillation amplitude. Given its advantages of high operation frequency, superior spatial resolution, and phase-sensitive detection of piezoelectric potential, the TMIM is expected to play a major role in future studies of acousto-electronic systems.

To summarize, we report the fabrication of suspended LiNbO$_3$ acoustic waveguides and the visualization of slanted power flow of acoustic waves on such phononic devices. Combining finite-element modeling and transmission-mode microwave microscopy, we are able to directly measure the power flow angle, which is in good agreement with the calculated value. FFT analysis of the two orthogonal microwave images yields crucial information on the forward-propagating and reflected waves, from which the propagation loss can be evaluated. Our work demonstrates the exquisite sensitivity and high resolution of the TMIM technique in characterizing phononic circuits, which may find widespread applications in electromechanics, optomechanics, and quantum science and engineering.

## ACKNOWLEDGMENTS

The TMIM work was supported by NSF Division of Materials Research Grant DMR-2004536 and Welch Foundation Grant F-1814. The phononic device fabrication work was supported by DARPA Microsystems Technology Office (MTO) Near Zero Power RF and Sensor Operations (N-ZERO) project programs.

## DATA AVAILABILITY

The data that support the findings of this study are available from the corresponding author upon reasonable request.